\documentclass{tADP2e}

\usepackage{epstopdf}
\usepackage{subfigure}

\theoremstyle{plain}

\theoremstyle{definition}

\theoremstyle{remark}

\newcommand{\ket}[1]{\ensuremath{\left|{#1}\right\rangle}}

\newcommand{\pare}[1]{\left(#1\right)}
\newcommand{\projsm}[2]{\vert#1\rangle\langle#2\vert}
\newcommand{\proj}[2]{\left\vert#1\rangle\langle#2\right\vert}

\begin{document}

\title{\textit{Digital-Analog Quantum Simulations with Superconducting Circuits}}

\author{
\name{Lucas Lamata\textsuperscript{a}$^{\ast}$\thanks{$^\ast$Corresponding author. Email: lucas.lamata@gmail.com}, Adrian Parra-Rodriguez\textsuperscript{a}, Mikel Sanz\textsuperscript{a}, and Enrique Solano\textsuperscript{a,b,c}}
\affil{\textsuperscript{a}Department of Physical Chemistry, University of the Basque Country UPV/EHU, Apartado 644, 48080 Bilbao, Spain \\
\textsuperscript{b}IKERBASQUE, Basque Foundation for Science, Maria Diaz de Haro 3, 48013 Bilbao, Spain \\
\textsuperscript{c}Department of Physics, Shanghai University, 200444 Shanghai, China}
\received{\today}
}

\maketitle

\begin{abstract}
Quantum simulations consist in the intentional reproduction of physical or unphysical models into another more controllable quantum system. Beyond establishing communication vessels between unconnected fields, they promise to solve complex problems which may be considered as intractable for classical computers. From a historic perspective, two independent approaches have been pursued, namely, digital and analog quantum simulations. The former usually provide universality and flexibility, while the latter allows for better scalability. Here, we review recent literature merging both paradigms in the context of superconducting circuits, yielding: digital-analog quantum simulations. In this manner, we aim at getting the best of both approaches in the most advanced quantum platform involving superconducting qubits and microwave transmission lines. The discussed merge of quantum simulation concepts, digital and analog, may open the possibility in the near future for outperforming classical computers in relevant problems, enabling the reach of a quantum advantage.
\end{abstract}

\begin{classcode} 03.67.Ac Quantum algorithms, protocols, and simulations; 85.25.Cp Josephson devices; 42.50.Pq Cavity quantum electrodynamics; micromasers
\end{classcode}


\newpage

{\abstractfont\centerline{\bfseries Index}\vspace{12pt}
\hbox to \textwidth{\hsize\textwidth\vbox{\hsize18pc
\hspace*{-12pt} {1.}    Introduction\\
\hspace*{7pt} {1.1.}  Quantum simulations\\
\hspace*{7pt} {1.2.}  Digital quantum simulations\\
\hspace*{7pt} {1.3.}  Analog quantum simulations\\
\hspace*{7pt} {1.4.}  Digital-analog quantum simulations\\
\hspace*{-12pt}{2.}    Techniques for digital quantum simulations\\
\hspace*{7pt} {2.1.}  Trotter-Suzuki expansion\\
\hspace*{7pt} {2.2.}  Jordan-Wigner and Bravyi-Kitaev transformations\\
\hspace*{7pt} {2.3.}  M{\o}lmer-S{\o}rensen gates and multiqubit gates\\
\hspace*{-12pt}{3.}    Digital quantum simulations\\
\hspace*{7pt} {3.1.}  Digital quantum simulation of spin models\\
\hspace*{7pt} {3.2.}  Digital quantum simulation of fermionic systems\\}
\hspace{-18pt}\vbox{\noindent\hsize18pc
\hspace*{-12pt}{4.}    Analog quantum simulations\\
\hspace*{7pt} {4.1.}  Analog quantum simulation of the quantum Rabi model\\
\hspace*{7pt} {4.2.}  Analog quantum simulation of Dirac physics\\
\hspace*{7pt} {4.3.}  Analog quantum simulation of Jaynes-Cummings and Rabi lattices\\
\hspace*{-12pt}{5.}    Digital-analog quantum simulations\\
\hspace*{7pt} {5.1.}  Digital-analog quantum simulation of the quantum Rabi model\\
\hspace*{7pt} {5.2.}  Digital-analog quantum simulation of the Dicke model\\
\hspace*{7pt} {5.3.}  Digital-analog quantum simulation of fermion-fermion scattering in quantum field theories\\
\hspace*{-16pt} {6.}    Perspectives towards digital-analog quantum computing \\
\hspace*{-12pt} {7.}    Conclusions \\
\hspace*{-12pt} {8.}    Funding \\
\hspace*{-12pt} {9.}    References \\
 }}}

\section{Introduction}

\subsection{Quantum simulations}

The field of superconducting circuits has dramatically advanced in the last years~\cite{Devoret13}. Nowadays, circuit quantum electrodynamics (cQED)~\cite{Wallraff04}, as well as, e.g., capacitively or inductively coupled superconducting qubits, are considered as potential scalable quantum platforms for quantum computing. Indeed, several milestones in quantum algorithms~\cite{Fedorov12} and foundations of quantum mechanics tests~\cite{Abdumalikov13} have been already performed. Single-qubit and two-qubit operations~\cite{Chow12}, entangled state preparation~\cite{Neeley10}, as well as protocols for fault tolerance and quantum error correction~\cite{Reed12, Barends14}, are some of the quantum technology tasks that can be faithfully performed. However, up to now, the most promising achievements attained with superconducting circuits are in the field of quantum simulations, due to their complexity and potential scalability.

A quantum simulator is in essence a controllable quantum platform whose aim is to mimic the dynamical or static properties of another, typically less controllable, quantum system. Originally, quantum simulations were proposed by Feynman~{\cite{Feynman82}, while some years later a rigorous mathematical analysis showing this idea was indeed feasible was produced~\cite{Lloyd96}. In the last years, several theoretical proposals for quantum simulations in different fields with superconducting circuits have been presented~\cite{Ripoll08, Tian10, Pritchett, Zhang13, Mei13, Ballester12, Pedernales13, Viehmann13, LasHeras, LasHerasFermion, GarciaAlvarez16, DiCandia15,Sweke16}, while five pioneering experiments on digital quantum simulations with superconducting circuits have been performed~\cite{digital exp1, digital exp2, digital exp3, Langford17, Kandala17}, showing the great potential of the field.
	
\begin{figure}[h]
	\centering
	\includegraphics[width=.9\linewidth]{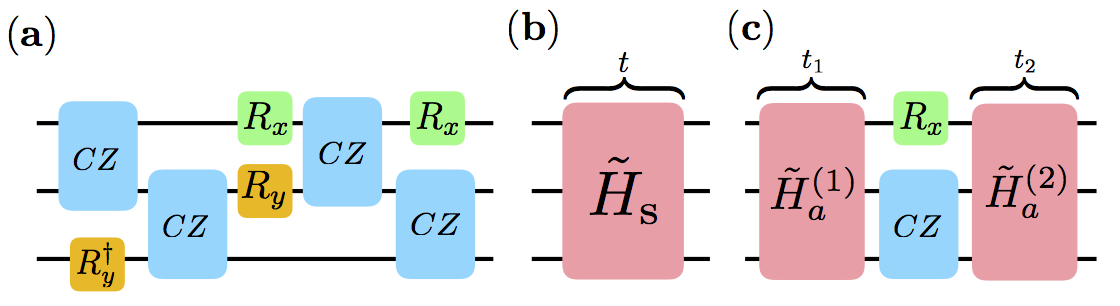}
	\caption{\label{fig:Abstract_DAQS} \textbf{Examples of digital, analog and digital-analog quantum simulation protocols.} ({\bf a}) A purely digital sequence of one and two qubit gates tries to reproduce a targeted quantum evolution with a generic Hamiltonian $H_s$. ({\bf b}) The analog simulator uses a controllable system whose evolution $\tilde{H}_s$ mimics almost one-to-one that of the simulated system $H_s$. ({\bf c}) A digital-analog quantum simulator combines analog blocks $\tilde{H}_{a}^{(i)}$ that naturally appear in the simulator with digital gates, e.g. entangling $CZ$-gates.}
\end{figure}
	
The field of quantum simulations can be roughly divided into digital quantum simulations, analog quantum simulations, and a more recent combination of both, namely, digital-analog quantum simulations, all of them represented in Fig.~\ref{fig:Abstract_DAQS}. While the purely digital approach is flexible and universal, the analog approach excels at robustness in the scope of scalability properties, allowing for the realization of simulations with a larger number of particles. A winning strategy already followed in the early ages of classical computers with the hybrid digital-analog computers is to combine the best of both worlds, flexibility and scalability, in a novel kind of digital-analog quantum simulator. In the latter, analog blocks provide a scalable structure by reducing the number of gates and hence the experimental error, while digital steps enhances the variety of possible interactions. We point out that this approach departs from a purely digital approach in the sense that it employs typically large analog blocks naturally appearing in each quantum platform, instead of decomposing the dynamics onto only single and two-qubit gates.

In this review article, we firstly introduce the different paradigms of quantum simulations which will be later on analyzed in the context of superconducting circuits: digital, in Sec. 1.2, analog, in Sec. 1.3, and digital-analog, in Sec. 1.4. In Sec. 2, we will introduce the basic techniques which are customarily employed for digital quantum simulations, namely, Trotter-Suzuki expansion, in Sec. 2.1, Jordan-Wigner and Bravyi-Kitaev transformations, in Sec. 2.2, and M{\o}lmer-S{\o}rensen gates and multiqubit gates, in Sec. 2.3. Then, we will review a selection of the literature in each of the three broad topics of quantum simulations. We will begin with the flexibility of the digital approach, in Sec. 3, following with the scalability of the analog techniques, in Sec. 4, and the convergence onto the mixed digital-analog paradigm, in Sec. 5. We believe this last kind of quantum simulator, that joins the universality of the digital techniques with the possibility of analog simulators to reach large sizes, may represent one of the most fruitful avenues to reach a quantum advantage. This will likely establish the field of digital-analog quantum simulations with superconducting circuits as a landmark on the way to outperforming classical computers and acquiring new and important knowledge about nature. Finally, in Sec. 6, inspired by the power of the merge between digital and analog viewpoints, we discuss possible perspectives for the digital-analog approach towards a universal digital-analog quantum computing framework.

\subsection{Digital quantum simulations}
The concept of digital quantum simulator was proposed by Lloyd in 1996~\cite{Lloyd96}, and is based in its most common formulation in a Trotter-Suzuki decomposition~\cite{Suzuki76} of the evolution operator of the simulated system, see Sec.~\ref{JordanWignerTrotter}. The idea is to decompose the evolution associated to a complex Hamiltonian or Lindbladians~\cite{DiCandia15,Sweke16} in terms of interactions implementable in the simulator, and particularly, in terms of single- and two-qubit gates (see Fig.~\ref{fig:Abstract_DAQS}.a). In this way, one can outperform the limitations of the simulating quantum platform, whose dynamics may be very different from the quantum system to be simulated. This is consequently a completely flexible approach which allows for quantum error correction and, finally, universality.

Besides theory proposals for digital quantum simulations, pioneering experiments have been performed both in trapped ions~\cite{Lanyon11,EstebanMartinez16} and in superconducting circuits~\cite{digital exp1,digital exp2,digital exp3}. Although they are still far from reaching quantum advantage, i.e. outperforming classical computers, these proposals and experiments have demonstrated the plethora of quantum systems which can be simulated. Indeed, the numbers of qubits and gates which can be experimentally employed, being significant, are still far from outperforming classical computers to solve interesting problems.

In Sec.~\ref{sec:DQS}, we will describe the proposals for digital quantum simulations of spin models~\cite{LasHeras} and fermionic systems~\cite{LasHerasFermion} with superconducting circuits, both of which have been carried out in the lab~\cite{digital exp1,digital exp2,digital exp3}.

\subsection{Analog quantum simulations}
Analog quantum simulations are other type of paradigm, whose aim is to mimic the dynamical or static properties of the simulated quantum model via reproducing its characteristics as close as possible in the simulating system for the whole continuous evolution (see Fig.~\ref{fig:Abstract_DAQS}.b). Therefore, analog quantum simulators are quantum platforms whose Hamiltonians are as similar as possible to the simulated systems. Some quantum systems show certain degree of tunability, such that the simulator Hamiltonian may approach the simulated one in different parameter regimes and conditions. The main appeal of analog quantum simulators is their scalability properties, since smaller errors allow for implementing larger quantum systems with these analog devices. Unfortunately, this approach shows in general less flexibility than a digital quantum simulator and they are single-purpose devices. 

In Sec.~\ref{sec:AQS}, we review the proposals and experiments for analog quantum simulators of the quantum Rabi model~\cite{Ballester12, Braumuller17}, Dirac physics~\cite{Dirac,Pedernales13}, as well as Jaynes-Cummings and Rabi lattices~\cite{CarusottoCiutiReview,Fisher1989,Hartmann2006,Hartmann2007,Greentree2006,Angelakis2007}.  It is worth to point out that analog quantum simulations of the quantum Rabi model have been proposed and recently realized in a different quantum platform, namely, trapped ions~\cite{PedernalesRabi,KihwanRabi,Puebla17,Puebla16}.

\subsection{Digital-analog quantum simulations}
In recent years, a novel approach to quantum simulations, which joins the best properties of digital and analog quantum simulators, has flourished, and it is called digital-analog quantum simulators (see Fig.~\ref{fig:Abstract_DAQS}.c). This new concept is based on devices employing large analog blocks, which provide scalability to the system, together with digital steps, which provide flexibility. This strategy, which mimics the strategy followed during the early ages of classical computers, will expectably allow us to outperform classical computers in solving relevant quantum models of significant complexity faster. In other words, we consider that this is the most promising avenue towards achieving quantum advantage without the need of quantum error correction protocols.

In Sec.~\ref{sec:DAQS}, we review some proposals for digital-analog quantum simulators, including the digital-analog simulation of the quantum Rabi model~\cite{MezzacapoRabi}, already carried out in an experiment~\cite{Langford17}, of the Dicke model~\cite{MezzacapoRabi,Lamata16}, and of fermion-fermion scattering in the context of quantum field theories~\cite{QFTcQED}, with superconducting circuits. However, this is just a summary of other relevant applications, such as in quantum biochemistry~\cite{GarciaAlvarez16}. Finally, in Sec.~\ref{sec:DAQC}, we propose the idea of digital-analog quantum computing as a theoretical framework to extend these ideas to a universal quantum computer.

\section{Techniques for digital quantum simulations}
\label{JordanWignerTrotter}
\subsection{Trotter-Suzuki expansion}
In many situations, a given quantum simulator cannot directly reproduce the dynamics of the model quantum system aimed for. Digital quantum simulations~\cite{Lloyd96} enable the possibility to engineer arbitrary quantum dynamics via decomposition of the original evolution in terms of efficiently implementable gates. This approach, known as the Lie-Trotter-Suzuki decomposition~\cite{Suzuki90}, consists in expanding the unitary evolution operator $e^{-iHt}$ associated with a Hamiltonian $H=\sum_k^M H_k$, in terms of a set of efficiently implementable gates, $e^{-iH_kt}$.  The Trotter expansion  reads $(\hbar=1)$
\begin{eqnarray}
e^{-iHt} \simeq \left(e^{-iH_1t/l} \cdots e^{-iH_Mt/l}\right)^l +\sum_{i<j}  \frac{[H_i,H_j]t^2}{2l}. \label{TrotterBasic}
\end{eqnarray} 
Here, $l$ is the total number of Trotter steps. By increasing the value of $l$, the digitized dynamics can be made as accurate as desired. As can be appreciated in Eq.~(\ref{TrotterBasic}), the largest error contribution in this approximation scales with $t^2/l$, as well as with the commutators of the Hamiltonians involved, which in turn are proportional to their norms or, e.g., coupling constants. Therefore, the longer the simulated time is, the larger the number of digital steps one needs to apply in order to achieve good fidelities.

\subsection{Jordan-Wigner and Bravyi-Kitaev transformations}
\label{sec:JW}

Fermions are quantum particles with semi-integer spin that obey Fermi-Dirac statistics. They are responsible for a wide variety of physical phenomena, involving material science, quantum chemistry, and high-energy physics. However, in order to universally simulate their properties via a digital quantum simulator, normally one has to map the fermionic behaviour onto the quantum simulator basic units, the qubits. There are different ways to achieve this and, here, we review two of the best known, namely, the Jordan-Wigner transformation~\cite{Jordan28} and the Bravyi-Kitaev transformation~\cite{Bravyi2002,Love15}.

The Jordan-Wigner (JW) transformation permits to establish a mapping between fermionic creation/annihilation operators and spin Pauli operators. In one spatial dimension this mapping preserves locality, i.e., a local fermion model maps onto a local spin one, while, for two or three spatial dimensions, a local fermionic Hamiltonian is mapped onto a nonlocal spin one. The JW mapping is given by $b_k^\dagger=I_N\otimes I_{N-1}\otimes ...\otimes\sigma_{k}^+\otimes\sigma_{k-1}^z\otimes ... \otimes\sigma_{1}^z$, and $b_k=(b_k^\dagger)^\dagger$, where $b_k(b_k^{\dagger})$ are fermionic annihilation (creation) operators and $\sigma_i^\alpha$ are the Pauli spin operators of the $i$th site, where $\sigma^+=(\sigma^x+i\sigma^y)/2$. The number of qubits which are necessary for implementing a single fermionic operator with this technique grows as $O(n)$, with $n$ the number of considered fermionic modes.

The JW transformation encodes the fermionic Hamiltonian in a local number basis, while the parity information, namely, the anticommutation-related phases, is nonlocally encoded. On the other hand, the Bravyi-Kitaev transformation employs a mixed encoding between number and parity bases, in such a way that they are both partially nonlocal. The advantage of using this approach is that the number of necessary qubit operations grows in this case logarithmically in the number of fermionic modes, $O(\log(n))$.  For further details of this technique, see Ref. \cite{Love15}.

\subsection{M{\o}lmer-S{\o}rensen gates and multiqubit gates}
\label{sec:MolmerSorensen}

The M{\o}lmer-S{\o}rensen gates (MS)~\cite{Molmer99} are the most commonly used multiqubit entangling operations in trapped ions. They allow for generating highly entangled states involving several ions in long chains, and they have been proven very useful for performing digital or digital-analog quantum simulations of fermionic models with trapped ions. The unitary operator which produces this gate is given by
$U_{S_z^2}=\exp[-i\pi/4\sum_{i<j}\sigma_i^z\sigma_j^z]$. By appropiate combination of two MS gates with a local gate acting on one qubit, arbitrary $k$-body spin Hamiltonian dynamics may be implemented,
\begin{eqnarray}
U=U_{S_z^2}U_{\sigma_y}(\phi)U^{\dagger}_{S_z^2}=\exp[i\phi\sigma_1^y\otimes\sigma_2^z\otimes\sigma_3^z\otimes...\otimes\sigma_k^z],\label{MSDecompos}
\end{eqnarray} 
where $U_{\sigma_y}(\phi)=\exp[-i\phi'\sigma_1^{y(x)}]$ for odd(even) $k$. The phase $\phi'$ depends as well on the number of qubits, i.e., $\phi'=\phi$ for $k=4n+1$, $\phi'=-\phi$ for $k=4n-1$, $\phi'=-\phi$ for $k=4n-2$, and $\phi'=\phi$ for $k=4n$, where $n$ corresponds to a positive integer. Via this unitary evolution and by introducing single-qubit operations, it is feasible to generate any tensor product of Pauli operators during an evolution phase given in terms of $\phi$. With this tool, one may implement arbitrary fermionic Hamiltonian dynamics in more than one spatial dimension, which is classically unfeasible.

Recently, variants of this protocol have been proposed for superconducting circuits via tunable couplings of transmon qubits to resonators~\cite{Mezzacapo14}. By appropriate combinations of detuned red and blue sideband interactions, these gates may be implemented in circuit QED with high fidelities.

\section{Digital quantum simulations}
\label{sec:DQS}
In this Section we review two of the main topics addressed in digital quantum simulations with superconducting circuits, namely, spin models~\cite{LasHeras} and fermionic systems~\cite{LasHerasFermion}.

\subsection{Digital quantum simulation of spin models}
We now review a protocol for implementing a Heisenberg or Ising dynamics with superconducting circuits, via digital techniques~\cite{LasHeras}. A detailed description of the circuit for this quantum simulation can be found in Ref.~\cite{digital exp2}.

{\it Heisenberg interaction.- }
The Heisenberg spin dynamics can be simulated with state-of-the-art technology in superconducting circuits via digital techniques. We consider an implementation composed of several transmon qubits which are coupled to a microwave resonator~\cite{Koch07},
\begin{eqnarray}
H^T=\omega_ra^{\dagger}a+\sum_{i=1}^N\Big{[}4E_{C,i}(n_i-n_{g,i})^2-E_{J,i}\cos\phi_i+2\beta_i eV_{\textrm{rms}}n_i(a+a^{\dagger})\Big{]}.\label{HT}
\end{eqnarray}
Here, $n_i$, $n_{g,i}$ and $\phi_i$ denote, respectively, the charge quanta on the superconducting island, the offset charge and the quantum flux of the $i$-th transmon qubit. The bosonic operators $a$($a^{\dagger}$) correspond to the resonator field, whose first mode frequency is $\omega_r$. $E_{J,i}=E_{J,i}^{\textrm{max}}|\cos(\pi\Phi_i/\Phi_0)|$ is the Josephson energy of the dc-SQUID loop embedded in the $i$-th qubit, while $E_{C,i}$ is the charging energy of the superconducting island. Also, $\beta_i$ are coupling coefficients due to circuit capacitances, $V_{\textrm{rms}}$ is the rms resonator voltage, and $e$ is the unit charge. Typical regimes for the transmon consider ratios $E_J/E_C\gtrsim20$. 

The Heisenberg interaction does not appear in cavity QED or circuit QED from first principles. Nevertheless, one can achieve it via a digital quantum simulation.
Here, we show that the interacting system of transmon and resonator described in Eq.~(\ref{HT}), can implement Heisenberg dynamics for $N$ qubits, which in the homogeneous case reads
\begin{equation}
H^{\textrm{H}}=\sum_{i=1}^{N-1}  J\left(\sigma^x_i \sigma^x_{i+1} +\sigma^y_i \sigma^y_{i+1} +\sigma^z_i \sigma^z_{i+1}\right).\label{XYZinhom}
\end{equation}
Here, $\sigma^j_i$, $j\in\{x,y,z\}$ are Pauli matrices which refer to the subspace spanned by the $i$-th transmon qubit. 
 We consider first the simplest two-qubit case. The XY exchange dynamics is the naturally appearing one in circuit QED, and can be directly achieved by dispersively coupling two transmon qubits to the same coplanar waveguide~\cite{Blais04,Majer07,Filipp11}, $H_{12}^{xy}=  J \left( {\sigma_1^+} {\sigma_2 ^-}+  {\sigma_1^-} {\sigma_2^+} \right) =   J/2  \left(\sigma^x_1 \sigma^x_2 +  \sigma^y_1 \sigma^y_2\right)$.
The exchange interaction can be then mapped via local rotations of the single qubits to the effective Hamiltonians 
$H_{12}^{xz} = R^x_{12}(\pi/4)H_{12}^{xy}R^{x \dagger}_{12}(\pi /4) = J/2  \left(\sigma^x_1 \sigma^x_2 +  \sigma^z_1 \sigma^z_2 \right)$ and $H_{12}^{yz}=R^{y}_{12}(\pi/4)H_{12}^{xy}R^{y \dagger}_{12}(\pi /4)= J/2  \left(\sigma^y_1 \sigma^y_2 +  \sigma^z_1 \sigma^z_2 \right)$. Here, $R_{12}^{x(y)}(\pi/4)=\exp[-i\pi/4(\sigma_1^{x(y)}+\sigma_2^{x(y)})]$ denotes a local rotation of first and second transmon qubits along the $x(y)$ axis.
The XYZ Heisenberg Hamiltonian $H_{12}^{xyz}$ can accordingly be performed combining these three interactions. Consequently, the total unitary evolution reads
\begin{eqnarray}
U_{12}^{\textrm{H}} (t)&=& e^{-i H_{12}^{xy} t} e^{-i H_{12}^{xz} t} e^{-i H_{12}^{yz} t}= e^{-i H^{\textrm{H}}_{12}t}.\label{eq8}
\end{eqnarray}
This unitary operator implements the dynamics of Eq.~(\ref{XYZinhom}) for just two qubits. Inhomogeneous couplings can also be achieved by accumulating different simulated phases for different discretized steps. We point out that, in the two qubit case, just a single Trotter step is needed to achieve a simulation with no digital error, due to the fact that $H_{12}^{xy}$, $H_{12}^{xz}$, and $H_{12}^{yz}$ commute. Therefore, in this case, the only error sources will appear from the accumulation of gate errors. Assuming two-qubit operations with an error of about $5\%$ and eight $\pi /4$ single qubit gates with errors of $1\%$, we obtain a total fidelity of the protocol of around $77\%$. We point out that these are just illustrative values and in many superconducting circuit experiments fidelities are nowadays better~\cite{Barends14, digital exp1, digital exp3}. Additionally, the total experimental time for a $\pi/4$ simulated XYZ phase will be of around $0.10$~$\mu$s, according to typical circuit QED gate times. 

\begin{figure}[h]
	\centering
	\includegraphics[width=.9\linewidth]{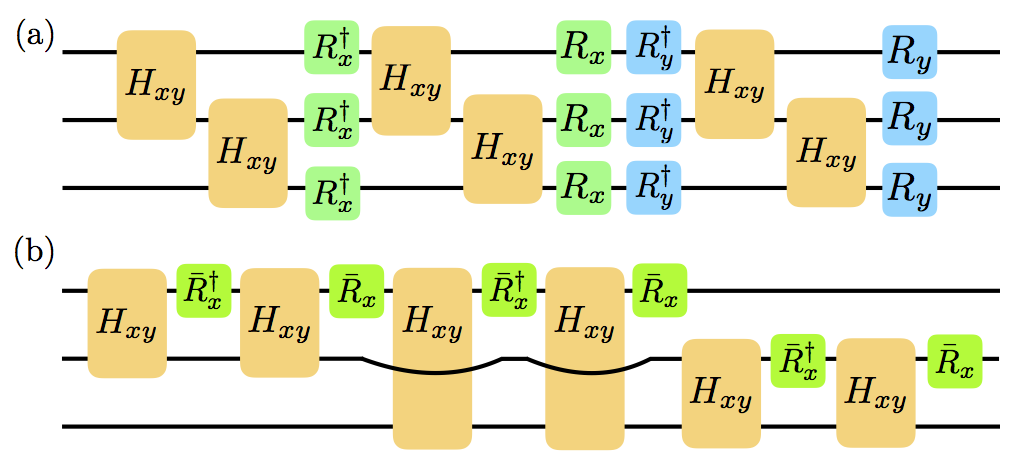}
	\caption{\label{fig:DQ_SpinModels_Qcircuit} {\bf Protocols for digital quantum simulations}. Trotter steps to simulate (a) the Heisenberg model and (b) the frustrated Ising model for three qubits. Here, $R_{x(y)}\equiv R^{x(y)}(\pi/4)$ and $R_{x(y)}\equiv \bar{R}^{x(y)}(\pi/2)$. We can exchange each $R$ matrix with its adjoint and the protocols will not get affected.}
\end{figure}

The case of three or more spins can be tackled in a similar way, but, due to the noncommutativity of the terms involved, in these cases it becomes necessary to employ Trotterization. In this situation, the Trotter step has to be repeated $l$ times following Eq.~(\ref{TrotterBasic}), to emulate the dynamics of Eq.~(\ref{XYZinhom}) for three qubits (see Fig.~\ref{fig:DQ_SpinModels_Qcircuit}.a). Each Trotter step has four single qubit rotations at different times and six two-qubit operations, achieving a step time of around $0.16$~$\mu$s, well below standard decoherence times for transmon qubits~\cite{Rigetti12}. 

{\it Ising interaction.- }
Now we consider an $N$-qubit Ising dynamics given by $J\sum_i\sigma_i^x\sigma_{i+1}^x$, with periodic boundary conditions. Analyzing a three site model is enough to evidence the effect of frustration in the model. The antiferromagnetic dynamics is inefficient with classical computers, while being efficient with quantum simulators~\cite{KimFrustration}. We study the isotropic antiferromagnetic Ising model among three spins, $H^{\textrm{I}}_{123} =  J \sum_{i<j}\sigma_i^x \sigma_j^x$, with $i,j=1,2,3$ and $J>0$. To simulate this model, one may apply a $\pi/2$ rotation to one qubit, in order to change the sign of the YY term in the XY Hamiltonian. This will produce an effective stepwise elimination of this term,
\begin{equation}
H_{12}^{x-y}={R}^{x}_1(\pi/2)H_{12}^{xy}{R}_1^{x\dagger}(\pi /2) = J \left(\sigma^x_1 \sigma^x_2 -  \sigma^y_1 \sigma^y_2 \right).
\end{equation}
Given that the terms in the Ising Hamiltonian commute, there is no Trotter error, and we can consider a single Trotter step, depicted in Fig.~\ref{fig:DQ_SpinModels_Qcircuit}.b. We estimate a protocol fidelity of about $64\%$ and an estimated time for the protocol of $0.18$~$\mu$s.                                                                                                                            

We can also consider adding a transverse magnetic field, which produces the Hamiltonian $H^{\textrm{IT}}_{123} =  J \sum_{i<j}\sigma^x_i \sigma^x_j + B \sum_i \sigma^y_i$. In this situation, the Hamiltonian terms do not commute anymore, such that we need to employ several Trotter steps to achieve high fidelities. The unitary dynamics of a single Trotter step in this case reads,
\begin{eqnarray}
U(t/l)=&& e^{-i H_{12}^{xy} t/l}e^{-i H_{12}^{x-y} t/l}e^{-i H_{13}^{xy}t/l} e^{-i H_{13}^{x-y}t/l}\nonumber\\
&&\times e^{-i H_{23}^{xy}t/l}e^{-i H_{23}^{x-y}t/l}e^{-i  Bt/l(\sigma^y_1+\sigma^y_2+\sigma^y_3)}\label{eq20}\\
&&= e^{-i  2Jt/l(\sigma^x_1 \sigma^x_2 + \sigma^x_1 \sigma^x_3 + \sigma^x_2 \sigma^x_3) }e^{-i  Bt/l(\sigma^y_1+\sigma^y_2+\sigma^y_3)}.\nonumber
\end{eqnarray}

Recently, an interesting scalable method based on genetic algorithms has been proposed to generate alternative quantum circuits to the aforementioned ones~\cite{LasHeras16}. Indeed, for the same number of resources (number of two-qubit gates), a circuit simulating Ising and Heisenberg models with smaller digital and experimental error was generated. This is a promising avenue for reducing errors in mid-sized quantum simulators without making use of quantum error correction.

\subsubsection{Experimental realizations}

This proposal was carried out in a circuit QED experiment with two qubits, which implemented both the Heisenberg and the Ising dynamics~\cite{digital exp2}, and a scheme of the experiment is depicted in Fig.~\ref{fig:DQ_SpinModels_circuit_ETH}. A more recent experiment employed Xmon superconducting qubits to perform a digitized adiabatic quantum evolution of non-stoquastic spin models, employing up to 9 qubits and more than 1000 quantum logic gates~\cite{digital exp3}.

\begin{figure}[h]
	\centering
	\includegraphics[width=.8\linewidth]{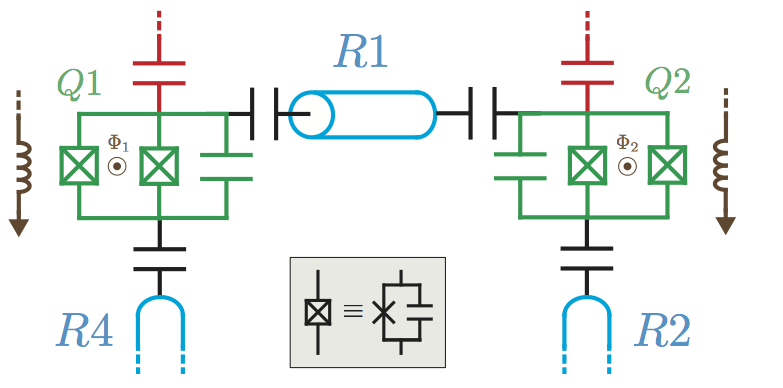}
	\caption{\label{fig:DQ_SpinModels_circuit_ETH} {\bf Partial circuital sketch for the experimental realization of DQS with transmon qubits in \cite{digital exp2}}. In the figure, two transmon qubits $Q1,\,Q2$ can be tuned through flux-biasing lines and are  connected through the transmission line resonator $R1$. The  square lattice chip of the experiment is completed with qubits $Q3$ and $Q4$ and resonator $R3$ (not drawn).}
\end{figure}

\subsection{Digital quantum simulation of fermionic systems}
\label{subsec:fermions}

\subsubsection{Fermi-Hubbard model: small lattices with pairwise interactions}

We now introduce a superconducting circuit encoding of the Fermi-Hubbard model, as an instance of a fermionic model with nearest-neighbour couplings, which hence employs only pairwise capacitive spin couplings in the one-dimensional case~\cite{LasHerasFermion}. A complete description of the experimental realization of this proposal can be found in Ref.~\cite{digital exp1}. For the sake of clarity, we focus on three fermionic modes, although these methods can straightforwardly be extended to arbitrary number of modes in higher spatial dimensions. These situations can in general be mapped onto multiqubit operations that can be always expanded into a polynomial number of two-qubit operations, as shown in Eq.~(\ref{MSDecompos}). In the last part of Sec.~\ref{subsec:fermions}, we analyze another cQED platform that employs resonators instead of direct qubit interactions to mediate the couplings~\cite{LasHerasFermion}.

The Fermi-Hubbard model is a solid state system describing interacting electrons in a lattice. The model describes the tradeoff between the electron kinetic energy, encoded in  hopping terms, with the Coulomb interaction, which is implemented by a nonlinear term. We first consider a small lattice realizable with current superconducting circuit technology. We study the Fermi-Hubbard-like model for three spinless fermionic modes with open boundary conditions,
\begin{equation}
H = -h \left(b^{\dag}_1 b_2 + b^{\dag}_2 b_1 + b^{\dag}_2 b_3 + b^{\dag}_3 b_2 \right) +U \left(b^{\dag}_1 b_1 b^{\dag}_2 b_2 + b^{\dag}_2 b_2 b^{\dag}_3 b_3 \right).
\label{Hub}
\end{equation}
Here, $b_m^{\dag}$ ($b_m$) are fermionic creation (annihilation) operators for site $m$.

We will employ the Jordan-Wigner transformation to map the fermionic operators onto tensor products of Pauli operators. We will show below that the latter can be efficiently realized in superconducting circuits. The Jordan-Wigner mapping for this case reads,
\begin{eqnarray}
b^{\dag}_1 &=& \mathbb{I} \otimes \mathbb{I} \otimes \sigma^{+} , \nonumber \\
b^{\dag}_2 &=& \mathbb{I} \otimes  \sigma^{+} \otimes \sigma^{z} , \nonumber \\
b^{\dag}_3 &=&  \sigma^{+} \otimes \sigma^{z} \otimes \sigma^{z}.
\end{eqnarray}
We then rewrite the Hamiltonian in Eq.~(\ref{Hub}) in terms of spin-$1/2$ Pauli operators, 
\begin{eqnarray}
H & = & \frac{h}{2} \left(\mathbb{I} \otimes \sigma^{x} \otimes \sigma^{x} + \mathbb{I} \otimes \sigma^{y} \otimes \sigma^{y} + \sigma^{x} \otimes \sigma^{x} \otimes \mathbb{I} + \sigma^{y} \otimes \sigma^{y} \otimes \mathbb{I} \right) \nonumber \\
&& + \frac{U}{4} \big(\mathbb{I} \otimes \sigma^{z} \otimes \sigma^{z} + \mathbb{I} \otimes \sigma^{z} \otimes \mathbb{I} + \mathbb{I} \otimes \mathbb{I} \otimes \sigma^{z} + \sigma^{z} \otimes \sigma^{z} \otimes \mathbb{I} \nonumber \\
&& + \sigma^{z} \otimes \mathbb{I} \otimes \mathbb{I} + \mathbb{I} \otimes \sigma^{z} \otimes \mathbb{I}  \big).
\label{Hubspin}
\end{eqnarray}
Here, each of the interactions can be implemented via digital decomposition using a specific gate sequence. We consider first the associated dynamics in terms of $\exp( -i \phi \sigma^{z} \otimes \sigma^{z} )$ interactions. These may be carried out in small steps of $CZ_{\phi}$ gates, where single-qubit gate and two-qubit gate fidelities of 99.92\% and to 99.4\%, respectively, have been already achieved~\cite{Barends14}. We can therefore employ the following expressions,
\begin{eqnarray}
\sigma^{x} \otimes \sigma^{x} &=& R_{y} (\pi/2)\sigma^{z} \otimes \sigma^{z} R_{y} (-\pi/2) , \nonumber \\
\sigma^{y} \otimes \sigma^{y} &=&  R_{x} (-\pi/2)\sigma^{z} \otimes \sigma^{z} R_{x} (\pi/2) ,
\end{eqnarray}
where $R_{j} (\theta) = \exp(-i\frac{\theta}{2} \sigma^{j})$ are local rotations along the $j$th axis of the Bloch sphere which act on both qubits.

The corresponding unitary operator associated with the Hamiltonian in Eq.~(\ref{Hubspin}) can be written in terms of $\exp( -i \phi \sigma^{z} \otimes \sigma^{z} )$ interactions. Moreover, these gates may be reordered in a more appropriate way in order to optimise the gate number,
\begin{eqnarray}
\exp(-iH t) &\approx& \bigg[ R^{'}_{y} (\pi/2) \exp\left(-i \frac{h}{2} \mathbb{I} \otimes \sigma^{z} \otimes \sigma^{z} \frac{t}{n}\right) R^{'}_{y} (-\pi/2) R_{y} (\pi/2) \nonumber \\
& & \times \exp\left(-i \frac{h}{2} \sigma^{z} \otimes \sigma^{z} \otimes \mathbb{I} \frac{t}{n}\right) R_{y} (-\pi/2) R^{'}_{x} (-\pi/2) \nonumber \\
& & \times  \exp\left(-i \frac{h}{2} \mathbb{I} \otimes \sigma^{z} \otimes \sigma^{z} \frac{t}{n}\right) R^{'}_{x} (\pi/2) R_{x} (-\pi/2) \nonumber \\
& & \times \exp\left(-i \frac{h}{2} \sigma^{z} \otimes \sigma^{z} \otimes \mathbb{I} \frac{t}{n}\right) R_{x} (\pi/2) \exp\left(-i \frac{U}{4} \mathbb{I} \otimes \sigma^{z} \otimes \sigma^{z} \frac{t}{n}\right)\nonumber \\
& & \times \exp\left(-i \frac{U}{2} \mathbb{I} \otimes \sigma^{z} \otimes \mathbb{I} \frac{t}{n}\right) \exp\left(-i \frac{U}{4} \mathbb{I} \otimes \mathbb{I} \otimes \sigma^{z} \frac{t}{n}\right) \nonumber \\
& & \times \exp\left(-i \frac{U}{4} \sigma^{z} \otimes \sigma^{z} \otimes \mathbb{I} \frac{t}{n}\right) \exp\left(-i \frac{U}{4} \sigma^{z} \otimes \mathbb{I} \otimes \mathbb{I} \frac{t}{n}\right) \bigg]^{n},
\end{eqnarray}
where we employ the prime in the rotation to differentiate between gates applied on different qubits, given that $R_{i}$ acts on first and second qubits, while $R^{'}_{i}$ acts on second and third.

The $\exp( -i \phi \sigma^{z} \otimes \sigma^{z} )$ operator can be carried out with optimized $CZ_{\phi}$ gates,
\[\exp\left(-i\frac{\phi}{2} \sigma^{z} \otimes \sigma^{z}\right) =
 \left( \begin{array}{cccc}
1 & 0 & 0 & 0 \\
 0 & e^{i\phi} & 0 & 0 \\
 0 & 0 & e^{i\phi} & 0 \\
 0 & 0 & 0 & 1
\end{array} \right).\]

\subsubsection{Large lattices and multiqubit gates mediated via a resonator}

Digital-analog quantum simulations of fermionic and bosonic systems, including quantum chemistry problems, have been already proposed in trapped ions~\cite{Casanova12, Mezzacapo12, Lamata14,Yung13}. In these articles, the use of multiqubit interactions mediated by a quantum bus, with additional digital decompositions, which have been recently implemented in ion-trap experiments~\cite{Muller11, Barreiro11}, allow for the implementation of arbitrary fermionic models.
Many current superconducting circuit systems are composed of superconducting qubits  and coplanar wave resonators~\cite{Devoret13}. A resonator, with its nonlocal character, can allow for the efficient implementation of the dynamics of two-dimensional and three-dimensional fermionic systems.

Recently, the realization of multiqubit gates via tunable transmon-resonator couplings has been analyzed~\cite{Mezzacapo14}. These interactions allow one for the implementation of multipartite entanglement~\cite{Mlyanek}, topological models~\cite{Kitaev}, and also the quantum simulation of fermionic systems. As shown in Sec.~\ref{sec:MolmerSorensen}, combining two multiqubit operations and a single-qubit rotation, the evolution operator associated with a nonlocal Hamiltonian made of a tensor product of spin operators can be built. Thus, via Jordan-Wigner transformation, introduced in Sec.~\ref{sec:JW}, any fermionic Hamiltonian can be digitally decomposed in terms of Trotter steps with the described multiqubit gate blocks. This allows one for the efficient implementation of fermionic dynamics in arbitrary spatial dimensions.

\subsubsection{Experimental realization}

Recently, the proposal for digital quantum simulations of fermionic models with superconducting circuits was carried out in the lab, in an implementation of a four-mode Hubbard model with four qubits and more than 300 quantum logic gates~\cite{digital exp1}.

\section{Analog quantum simulations}
\label{sec:AQS}
In this Section, we review three examples of analog quantum simulations with superconducting circuits, namely, the quantum Rabi model~\cite{Ballester12}, Dirac physics~\cite{Dirac,Pedernales13}, as well as Jaynes-Cummings and Rabi lattices~\cite{CarusottoCiutiReview,Fisher1989,Hartmann2006,Hartmann2007,Greentree2006,Angelakis2007}.

\subsection{Analog quantum simulation of the quantum Rabi model}

In this section, we review a protocol for the analog quantum simulation of the quantum Rabi model in ultrastrong coupling (USC) and deep strong coupling (DSC) regimes in circuit QED, using only a cavity-qubit system in the strong coupling regime~\cite{Ballester12}. A complete analysis of the experiment related to this proposal can be found in Ref.~\cite{Braumuller17}. This analysis employs an orthogonal two-tone driving to the qubit. Via this protocol, one will be able to access the regimes of USC ($0.1\lesssim \! g/ \omega \lesssim \! 1$, with $g/ \omega$ the ratio of the coupling over the frequency of the resonator) and DSC~\cite{DSC} ($g/ \omega\gtrsim \! 1$). This will enable the implementation of quantum simulators \cite{ReviewQS} for a wide variety of regimes of light-matter coupling \cite{Braak, Rossatto} in platforms where they are not available from first principles. This includes, among others, the simulation of relativistic quantum physics and Dirac equation, the Dicke model, and the Jahn-Teller instability~\cite{Duty}. We review this protocol in the framework of circuit QED, although it can also be carried out in microwave cavity QED~\cite{Haroche,Sol} and other platforms.

The model we consider is composed of a superconducting qubit, e.g., flux or transmon, strongly coupled to a coplanar microwave resonator mode. At the qubit degeneracy point, the Hamiltonian reads \cite{Blais04}
\begin{eqnarray}
{\cal H} =  \frac{\hbar \omega_q}{2} \sigma_z + \hbar  \omega a^\dag a  -\hbar  g    \sigma_x  \pare{a + a^\dag} \label{HamilDiag}, 
\end{eqnarray}
where $\omega$ and $\omega_q$  are the bosonic mode and qubit frequencies, while $g$ is the qubit-cavity coupling strength. Moreover, $a$($a^\dag$) denote the annihilation(creation) operators of the photonic field mode, whereas $\sigma_x = \sigma^\dag + \sigma = \projsm{e}{g}+ \projsm{g}{e}$, $\sigma_z = \projsm{e}{e}-\projsm{g}{g}$, where ${\ket{g},\ket{e}}$ denote the ground and excited qubit eigenstates. In standard circuit QED implementations, one can apply the rotating-wave approximation (RWA) to this Hamiltonian. Provided that~\cite{Zueco} $\{|\omega-\omega_q|, g\} \ll\omega+\omega_q$, one can simplify the interaction to
\begin{eqnarray}
{\cal H} &=& \frac{\hbar \omega_q}{2} \sigma_z +\hbar  \omega a^\dag a -\hbar  g \pare{\sigma^\dag a + \sigma a^\dag} ,\label{HamilRWA}
\end{eqnarray}
which has the form of the Jaynes-Cummings model of quantum optics. Notice that the Hamiltonian in Eq.~(\ref{HamilRWA}) has not counterrotating terms, such that the number of excitations is conserved.

We assume now that the qubit is addressed orthogonally by two classical drivings. The Hamiltonian of this system can be expressed as
\begin{eqnarray}
  {\cal H} &=&  \frac{\hbar \omega_q}{2} \sigma_z +\hbar  \omega a^\dag a -\hbar  g \pare{\sigma^\dag a + \sigma a^\dag}  \nonumber \\  & & \hspace{-13mm} - \hbar \Omega_1  \pare{ e^{i \omega_1 t} \sigma + e^{-i \omega_1 t} \sigma^\dag} - \hbar  \Omega_2 \pare{ e^{i \omega_2 t} \sigma + e^{-i \omega_2 t} \sigma^\dag}, \label{HamilDriv}
\end{eqnarray}
where $\omega_j$ and $\Omega_j$ denote the frequency and amplitude of the $j-$th driving pulse. We point out that the orthogonal drivings address the qubit similarly to the way the resonator bosonic field does. In order to achieve (\ref{HamilDriv}), we have not only assumed the RWA applied to the qubit-resonator interaction term, but also in the case of both classical drivings. 

The following step is to express (\ref{HamilDriv}) in the rotating frame at the first driving frequency, $\omega_1$, resulting in
\begin{eqnarray}
{\cal H}^{L_1} &=&\hbar  \frac{\omega_q-\omega_1}{2} \sigma_z +\hbar  (\omega-\omega_1) a^\dag a - \hbar g  \pare{\sigma^\dag a +\sigma a^\dag} \nonumber \\ & &  \hspace{-13mm} - \hbar \Omega_1  \pare{ \sigma + \sigma^\dag } - \hbar  \Omega_2   \pare{ e^{i (\omega_2-\omega_1) t} \sigma + e^{-i (\omega_2-\omega_1) t} \sigma^\dag } .
\end{eqnarray}
This permits to map the original first classical driving into a time-independent term ${\cal H}_0^{L_1} = - \hbar \Omega_1  \pare{ \sigma + \sigma^\dag} $. We also consider this term to be the most important one and we treat the others in a perturbative fashion, going into the rotating frame with respect to ${\cal H}_0^{L_1} $, ${\cal H}^{I} (t) = e^{i {\cal H}_{0}^{L_1} t/\hbar } \pare{{\cal H}^{L_1}   - {\cal H}_0^{L_1} }    e^{-i {\cal H}_{0}^{L_1} t/\hbar } $. Employing the rotated basis for the spin $\ket{\pm} = \pare{\ket{g} \pm \ket{e} }/\sqrt2$, we have
\begin{eqnarray}
{\cal H}^{I} (t)  \!  &=&  \!  -\hbar  \frac{\omega_q-\omega_1}{2}  \!\!  \pare{ e^{-i 2  \Omega_1  t}  \! \proj{+}{-}   \! +  \!   {\rm H.c.}} \!  + \hbar (\omega-\omega_1) a^\dag a \nonumber  \\
  & -& \frac{\hbar g }{2} \left( \left\{  \proj{+}{+} - \proj{-}{-} + e^{-i 2  \Omega_1 t} \proj{+}{-} \right. \right. \nonumber \\
 & & -  \left.\left. e^{i 2  \Omega_1  t}\proj{-}{+}  \right\} a + {\rm H.c.} \right) \nonumber \\  
  &-&  \frac{\hbar \Omega_2 }{2} \left(   \left\{  \proj{+}{+} - \proj{-}{-} - e^{-i 2  \Omega_1  t} \proj{+}{-}  \right. \right. \nonumber \\ & & \left. \left. + e^{i 2 \Omega_1  t}\proj{-}{+}  \right\} e^{i (\omega_2-\omega_1) t} + {\rm H.c.} \right). \label{HI1}
\end{eqnarray}
Adjusting the external driving parameters according to $\omega_1-\omega_2=2 \Omega_1 ,$ one can select the resonant terms in the resulting time-dependent Hamiltonian. Accordingly, in case we have a first driving, $ \Omega_1$, relatively strong, we can approximate the above expression by an effective Hamiltonian which is time-independent, and reads
\begin{eqnarray}
{\cal H}_{\rm eff}
=  \hbar (\omega-\omega_1) a^\dag a + \frac{\hbar\Omega_2}{2} \sigma_z   -  \frac{\hbar g}{2} \sigma_x \pare{a+a^\dag} .  \label{HamilEff}
\end{eqnarray}
We point out that the original Hamiltonian (\ref{HamilDiag}) and (\ref{HamilEff}) have the same form. While the coupling $g$ value is fixed in (\ref{HamilEff}), one can still adjust the effective parameters via tuning frequencies and amplitudes of the drivings. If one can achieve values according to $\Omega_2  \sim (\omega-\omega_1) \sim g/2$, the original system dynamics will perform a simulation of a qubit-resonator coupling with a relative strength outperforming the SC regime, and reaching USC/DSC regimes. This may be quantified via the ratio $g_{\rm eff} / \omega_{\rm eff}$, where $g_{\rm eff} \equiv g/2$ and $ \omega_{\rm eff}\equiv\omega-\omega_1$.

\subsection{Analog quantum simulation of Dirac physics}

In this section, we review an analog quantum simulation of the Dirac equation with circuit QED~\cite{Pedernales13}. In order to achieve this, we need a superconducting qubit, such as a flux qubit~\cite{flux}, at the degeneracy point, which is strongly coupled to a microwave field mode of a coplanar wave resonator. The corresponding interaction is well described by the Jaynes-Cummings dynamics (JCM) \cite{JC,Blais04,Wallraff04}. Moreover, we address the qubit via three additional classical microwave drivings, two of them transversal to the resonator~\cite{Ballester12}, which couple only to the qubit, and another one, longitudinal, coupling to the resonator mode. The corresponding time-dependent Hamiltonian reads
\begin{eqnarray}
  {\cal H} &=&  \frac{\hbar \omega_q}{2} \sigma_z +\hbar  \omega a^\dag a -\hbar  g \pare{\sigma^\dag a + \sigma a^\dag}   - \hbar \Omega  \pare{ e^{ i \pare{ \omega t +\varphi} } \sigma + e^{-i \pare{ \omega t +\varphi } } \sigma^\dag}  \nonumber \\ 
  & &  - \hbar  \lambda   \pare{ e^{i  \pare{ \nu t +\varphi }} \sigma + e^{-i  \pare{ \nu t +\varphi } } \sigma^\dag} + \hbar \xi \pare{  e^{i \omega t} a + e^{-i \omega t} a^\dag  }  , \label{HamilDriv2}
\end{eqnarray}
where $\sigma_y=i \sigma - i \sigma^\dag=i \proj{g}{e} - i \proj{e}{g}$ and $\sigma_z= \proj{e}{e} -  \proj{g}{g}$, with $\ket{g}$ and $\ket{e}$ the qubit ground and excited states. Here, $\omega_q$ and $\omega$ denote the qubit and photon frequencies, while $g$ is the coupling strength. Moreover, the two orthogonal driving pulses have respectively real amplitudes $\Omega$ and $\lambda$, phase $\varphi$, as well as frequencies $\omega$ and $\nu$. Additionally, the longitudinal driving pulse is described by an amplitude $\xi$ and a frequency $\omega$. We point out that so far we choose two of the classical pulses to be resonant with the resonator microwave mode, while the other amplitudes and frequencies will be later fixed. We additionally consider that $\omega_q=\omega$, i.e., qubit and resonator microwave field are resonantly coupled.

This protocol is composed of two subsequent transformations. First, the Hamiltonian in Eq.~(\ref{HamilDriv2}) is simplified by means of the rotating frame with the resonator frequency $\omega$,
\begin{eqnarray}
{\cal H}^{L_1} &=&  - \hbar g  \pare{\sigma^\dag a +\sigma a^\dag}  - \hbar \Omega \pare{ e^{ i \varphi } \sigma + e^{-i \varphi } \sigma^\dag} + \hbar \xi \pare{ a + a^\dag}    \nonumber \\ & & - \hbar  \lambda   \pare{ e^{i \left[ (\nu-\omega) t + \varphi \right] } \sigma + e^{-i \left[ (\nu-\omega) t  +  \varphi \right]  } \sigma^\dag } . \label{HL1}
\end{eqnarray}

Second, we map this Hamiltonian into another rotating frame with respect to ${\cal H}_0^{L_1} = - \hbar \Omega   \pare{ e^{ i \varphi } \sigma + e^{-i \varphi } \sigma^\dag}$. The corresponding Hamiltonian reads
\begin{eqnarray}
{\cal H}^{I}  \!  &=&  \! -  \frac{\hbar g }{2} \left( \left\{  \proj{+}{+} - \proj{-}{-} + e^{-i 2  \Omega t} \proj{+}{-} \right. \right. \nonumber \\
 & & -  \left.\left. e^{i 2  \Omega  t}\proj{-}{+}  \right\} e^{i \varphi} a + {\rm H.c.} \right) \nonumber \\  
  &-&  \frac{\hbar \lambda}{2} \left(   \left\{  \proj{+}{+} - \proj{-}{-} - e^{-i 2  \Omega  t} \proj{+}{-}  \right. \right. \nonumber \\ & & \left. \left. + e^{i 2 \Omega  t}\proj{-}{+}  \right\} e^{i (\nu-\omega) t} + {\rm H.c.} \right)   + \hbar \xi \pare{ a + a^\dag}   , \label{HI1_2}
\end{eqnarray}
where we have considered the alternative qubit basis $\ket{\pm} = \pare{\ket{g} \pm e^{-i \varphi} \ket{e} }/\sqrt2$. To further simplify this expression, we fix $\omega-\nu=2 \Omega ,$ while assuming the first driving amplitude, $\Omega$, to be large when compared to the other frequencies appearing in Eq.~(\ref{HI1_2}). Therefore, one can apply the rotating-wave approximation (RWA), producing the Hamiltonian
\begin{eqnarray}
{\cal H}_{\rm eff}
=  \frac{\hbar\lambda}{2} \sigma_z   +  \frac{\hbar g}{\sqrt2} \sigma_y \hat{p} + \hbar \xi \sqrt2 \, \hat{x} ,  \label{HamilEff2}
\end{eqnarray}
where $\varphi=\pi/2$ and we consider typical electromagnetic field quadratures, i.e., dimensionless $\hat{x} =(a+a^\dag)/\sqrt2$, $\hat{p} =-i(a-a^\dag)/\sqrt2$, with the corresponding commutation relation $\left[ \hat{x} , \hat{p} \right] = i$. We point out that $\Omega$ is not present in the effective Hamiltonian. This is due to the fact that the Hamiltonian was derived in a rotating frame with $\Omega$ acting as a dominant frequency and a large value as compared to the other ones.

The dynamics of the analyzed system coincides with that of the 1+1 Dirac equation, where $\hbar \lambda / 2$ and $\hbar g/\sqrt2$ are related to the mass and the vacuum speed of light, respectively. Moreover, the dynamics has an external potential $\hat{\Phi} = \hbar \xi \sqrt2 \, \hat{x}$ that depends linearly on the particle position. With this simulated dynamics, we can implement several ranges of physical regimes, from the ultrarrelativistic to the nonrelativistic ones. In this proposal, while the simulated Dirac mass is proportional to the $\lambda$ amplitude of the weak orthogonal driving, the linear potential strength is related to the longitudinal drive amplitude, $\xi$. We point out that a simulated massless particle corresponds to $\lambda = 0$ and $\nu = 0$, such that $\omega = 2 \Omega$ in Eq.~(\ref{HI1_2}). 

\subsection{Analog quantum simulation of Jaynes-Cummings and Rabi lattices}

Superconducting circuit technology offers unprecedented possibilities for control, tunability of physical parameters and scalability. Therefore, circuit QED is a leading platform to analyze many-body states of light \cite{CarusottoCiutiReview} via analog quantum simulations of the Bose-Hubbard (BH) dynamics \cite{Fisher1989,Hartmann2006,Hartmann2007}, the Jaynes-Cummings-Hubbard (JCH) dynamics \cite{Greentree2006,Angelakis2007}, as well as fractional quantum Hall effect with synthetic gauge fields \cite{Nunnenkamp2011,Angelakis2008,Koch2010,Hafezi2014,Grusdt2014, White2012}, spin models \cite{Cho2008,Kay2008}, and the Kagome lattice for bosons \cite{Houck2012,Underwood2012,Hamed2014}. 

In an analog quantum simulation of the Jaynes-Cummings-Hubbard dynamics,
\begin{equation}
H_{\rm JCH} = \sum^{N}_{j=1}\omega_0\sigma^{+}_{j}\sigma^{-}_{j} + \sum^{N}_{j=1}\omega a_{j}^{\dag}a_{j}+J\sum^{N}_{\langle i j \rangle}(a_{i}a^{\dag}_{j}+ \rm{H. c.}),
\end{equation}
the complete system consists of unit cells that may be composed of a transmon qubit coupled to a coplanar wave resonator \cite{SchmidtKoch2013}, being its dynamics described by the JC model. Moreover, to connect neighboring cells one may employ capacitive couplings between half-wave cavities lying in a linear array. We point out that a two-site JCH dynamics has been already performed in the lab \cite{Raftery2015}. Moreover, other complex geometries \cite{Nunnenkamp2011} may be achieved as well, establishing a significant step forward~\cite{HarocheRaymond}.

One of the prominent features of circuit QED is that it allows us to design nonlinear couplings between resonators through elements as Josephson junctions. More specifically, the cavity-cavity interaction which is mediated by SQUID setups could be a significant possibility for simulating the Bose-Hubbard evolution with attractive interactions \cite{Hartmann2010,Hartmann2013}, as well as the complete Bose-Hubbard and extended models \cite{Dimitris_Unpublished}. A relevant experiment already performed in a circuit QED setup is the Bose-Hubbard dimer \cite{Eichler2014}. This proposal and the experimental implementation in Ref. \cite{Raftery2015} motivate further theoretical analysis in order to simulate many-body states of matter coupled to light. Furthermore, the controlled dissipative evolution of many-body states of light \cite{DGAngelakis2009, DGAngelakis2010, Carusotto2009,MJHartmann2010,Grujic2012,Grujic2013,Schetakis2013}, as well as polariton phenomena in quantum information \cite{DGAngelakis2008, Kyoseva2010} may be additionally simulated with current circuit QED technologies. The interactions which appear in circuit QED permit to analyze a diversity of condensed matter models. One example is given by out-of-equilibrium dynamics of nonlinear cavity arrays for implementing photonic solid phases, as described by 
\begin{equation}
H = \sum_{i}[-\delta a^{\dag}_ia_i+\Omega(a_i+a^{\dag}_i)] - J\sum_{\langle i,j\rangle}(a_ia^{\dag}_j+{\rm H.c.}) 
+ U\sum_{i}n_i(n_i-1) + V\sum_{\langle i,j\rangle}n_{i}n_{j},
\end{equation}
which displays Bose-Hubbard interactions and Kerr nonlinearities among nearest neighbours \cite{Fazio2013}. The latter is provided by the Josephson nonlinear element in the SQUID loop. 

Circuit QED platforms allow as well to analyze two-dimensional lattices of interacting cavities \cite{Houck2012,Underwood2012}, allowing for the realization of a Kagome lattice for bosons. This also permits applying novel numerical methods such as projected entangled-pair states (PEPS) \cite{Hamed2014,Cirac2008,PerezGarcia2010}, with the motivation of analyzing the tradeoff between light and matter couplings, and discovering new many-body states of light.

\section{Digital-analog quantum simulations}
\label{sec:DAQS}
In this Section, we review three examples of digital-analog quantum simulations with superconducting circuits, namely, the quantum Rabi model~\cite{MezzacapoRabi}, the Dicke model~\cite{MezzacapoRabi,Lamata16}, and fermion-fermion scattering in the context of quantum field theories~\cite{QFTcQED}.

\subsection{Digital-analog quantum simulation of the quantum Rabi model}

We describe here the digital-analog quantum simulation of the quantum Rabi model with circuit QED~\cite{MezzacapoRabi}. The experimental realization of this proposal is fully described in Ref.~\cite{Langford17}. We begin considering the standard circuit QED setup comprising a transmon qubit~\cite{Koch07}, and a coplanar wave resonator. The Hamiltonian describing the system is~\cite{Blais04} 
\begin{equation}
H=\omega_r a^{\dagger}a +\frac{\omega_q}{2}\sigma^z +g(a^{\dagger}\sigma^-+a\sigma^+),\label{QubitResHam}
\end{equation}
where $\omega_q$ and $\omega_r$ are the qubit and resonator frequencies, $g$ is the qubit-resonator coupling strength, $a^{\dagger}$$(a)$ is the creation(annihilation) bosonic field operator for the resonator microwave mode, while $\sigma^{\pm}$ are the raising and lowering spin operators for the qubit. 
The capacitive coupling in Eq.~(\ref{QubitResHam}) assumes a two-level approximation of the transmon qubit, given that the coupling $g$ is typically much weaker than other transition frequencies of the system. Larger capacitive couplings may produce unintended transitions to higher-lying levels of the transmon, such that the coupling cannot be made arbitrarily large. On the other hand, analog quantum simulator methods based on composite pulses~\cite{Ballester12,Pedernales13} may produce an increase of the $g/\omega$ ratio, populating further the resonator, as shown in Sec.~\ref{sec:AQS}. Here, we introduce another approach, namely, a digital-analog quantum simulator, which may implement the dynamics of the quantum Rabi Hamiltonian,
\begin{equation}
H_R=\omega^R_r a^{\dagger}a +\frac{\omega^R_q}{2}\sigma^z +g^R\sigma^x(a^{\dagger}+a) ,\label{RabiHam}
\end{equation}
by encoding it in a circuit QED setup provided with a Jaynes-Cummings interaction, as in Eq.~(\ref{QubitResHam}). 

The quantum Rabi Hamiltonian in Eq.~(\ref{RabiHam}) can be expressed as sum of two terms, $H_R=H_1+H_2$, where
\begin{eqnarray}
&&H_1=\frac{\omega^R_r}{2} a^{\dagger}a +\frac{\omega^1_q}{2}\sigma^z +g(a^{\dagger}\sigma^-+a\sigma^+) , \nonumber \\
&&H_2=\frac{\omega^R_r}{2} a^{\dagger}a -\frac{\omega^2_q}{2}\sigma^z +g(a^{\dagger}\sigma^++a\sigma^-) ,
\label{Ham12} 
\end{eqnarray}
and the qubit transition frequency in each of the two steps is defined such that $\omega_q^1-\omega_q^2=\omega^R_q$. These two dynamics can be carried out in standard circuit QED setups with fast control of the qubit frequency. Beginning with the qubit-resonator interaction in Eq.~(\ref{QubitResHam}), one can transform into a frame rotating at $\tilde{\omega}$, in which the resulting effective interaction Hamiltonian reads 
\begin{equation}
\tilde{H}=\tilde{\Delta}_ra^{\dagger}a+\tilde{\Delta}_q\sigma^z+g(a^{\dagger}\sigma^-+a\sigma^+),\label{IntHam}
\end{equation}  
with $\tilde{\Delta}_r=(\omega_r-\tilde{\omega})$ and $\tilde{\Delta}_q=\left(\omega_q-\tilde{\omega}\right)/2$. Accordingly, Eq.~(\ref{IntHam}) matches $H_1$, with a proper relabeling of the coefficients.
The anti-Jaynes Cummings term $H_2$ can be performed digitally by applying a single-qubit rotation to $\tilde{H}$ with a different detuning for the qubit frequency,
\begin{equation}
e^{-i \pi\sigma^x/2}\tilde{H}e^{i \pi\sigma^x/2}=\tilde{\Delta}_ra^{\dagger}a-\tilde{\Delta}_q\sigma^z+g(a^{\dagger}\sigma^++a\sigma^-).\label{RotHam}
\end{equation}
By judiciously choosing different resonator-qubit detunings for the two steps, $\tilde{\Delta}^1_q$ in the first one and $\tilde{\Delta}^2_q$ in the second one, we will be able to implement the quantum Rabi dynamics, Eq.~(\ref{RabiHam}), through a digital quantum simulation~\cite{Lloyd96}, via interleaving the simulated evolutions. Typical quasiresonant Jaynes-Cummings interaction dynamics with different qubit frequencies are alternated with microwave drivings, in order to realize standard qubit flips~\cite{Blais04}. This protocol can be repeated several times, following digital simulation techniques, to achieve a better approximation of the quantum Rabi evolution.

The corresponding simulated Rabi parameters are obtained from the setup physical parameters in the following way. The simulated bosonic mode frequency relates to the resonator frequency detuning $\omega_r^R=2\tilde{\Delta}_r$, the qubit frequency is related to the transmon frequency in the two simulation steps, $\omega_q^R=\tilde{\Delta}_q^1-\tilde{\Delta}_q^2$, and the coupling qubit-resonator coincides between simulated and simulating systems, $g^R=g$. 

Taking into account that single-qubit operations take approximately~$\sim10$~ns, dozens of Trotter steps may be realized within the coherence time. We point out that, when realizing the protocol, one has to be careful not to populate higher levels of the transmon qubit. Considering typical transmon anharmonicities of $\alpha=-0.1$, and given large detunings with the microwave resonator, for a range of typical circuit QED parameters, it will not be populated.
Summarizing, one can simulate a wide variety of regimes by choosing different qubit frequency detunings and rotating pictures.

\subsection{Digital-analog quantum simulation of the Dicke model}

By combining multiple transmon qubits with a resonator, the previous digital-analog method to simulate the quantum Rabi model can be extended to implement the Dicke dynamics in a variety of parameter regimes and couplings~\cite{MezzacapoRabi,Lamata16}, 
\begin{equation}
H_D=\omega^R_r a^{\dagger}a +\sum_{j=1}^N\frac{\omega^R_q}{2}\sigma_j^z +\sum_{j=1}^Ng^R\sigma^x_j(a^{\dagger}+a).
\end{equation}
This digital-analog protocol can be implemented with polynomial resources via collective single-qubit rotations interleaved with collective Tavis-Cummings dynamics. Therefore, a significant feature of this approach is that the total simulation time does not scale with the system size $N$.
The Dicke model may thus be analyzed given sufficient coherence and low gate errors. We point out that this quantum simulation protocol is appropriate for circuit QED, since collective single-qubit addressing is feasible.

We point out that the Dirac equation dynamics appears as a particular case of the quantum Rabi model. For 1+1 dimensions, the Clifford algebra of the Dirac matrices reduces to that of Pauli operators, and the Dirac equation in a certain representation reads 
\begin{equation}
\label{Dirac equation 1+1}
i  \frac{d}{d t} \ket{\Psi} = (mc^2 \sigma_z + c p \sigma_x)\ket{\Psi},
\end{equation}
where $c$ is the speed of light, $m$ the particle mass, and $ p \propto (a - a^{\dagger}) / i$ the momentum operator. The Dirac Hamiltonian in Eq.~(\ref{Dirac equation 1+1}), $H_{\rm D}=mc^2 \sigma_z + c p \sigma_x$, presents the same structure as the quantum Rabi Hamiltonian, Eq.~(\ref{RabiHam}), for $\omega_r^R=0$. This is achieved by fixing $\tilde{\omega}=\omega_r$. The analogy is total by connecting $mc^2$ to $\omega^R_q/2$, $c$ to $g^R$, and the Dirac particle momentum to the microwave field quadrature, which may be measured with current technology~\cite{DiCandia}. This approach will permit the observation of the Dirac particle {\it Zitterbewegung}~\cite{Dirac,DiracExp}. 

Furthermore, an analysis of a digital-analog quantum simulator of generalized Dicke models in superconducting circuits has been recently put forward~\cite{Lamata16}.

\subsection{Digital-analog quantum simulation of fermion-fermion scattering in quantum field theories}

The most fundamental analysis of physical interactions relies on quantum field theories~\cite{Peskin}, where the analysis of interacting fermions and bosons is essential. Among other issues, these theories allow for the computation of cross sections, and descriptions of fermion-fermion scattering mediated by bosonic virtual and real excitations, self-energies of elementary particles, as well as boson vacuum polarization. In this section, we will review a model which describes a digital-analog quantum simulation of fermion-fermion scattering mediated by a bosonic mode~\cite{QFTcQED}, considering the following assumptions: (i)~one spatial and one temporal dimensions, (ii) scalar bosonic and fermionic particles, obeying the dynamics of the following Hamiltonian, 
\begin{eqnarray} \label{complex}
H = \int dp \ \omega_p (b^{\dag}_{p}b_{p} + d^{\dag}_{p}d_{p}) + \int dk \ \omega_k a^{\dag}_ka_k  
 +  \int dx \ \psi^{\dag}(x)\psi(x)A(x).
\end{eqnarray}
Here, $A(x)=i\int dk \ \lambda_k \sqrt{\omega_{k}}( a^{\dag}_k e^{-i k x} - a_k  e^{i k x} )/\sqrt{4\pi}$ is a bosonic field operator, where $\lambda_k$ are coupling constants, and $\psi(x) = \int dp\left( b_p  e^{i p x} +  d_p^{\dag} e^{-i p x} \right)/\sqrt{4\pi\omega_p}$ is a fermionic field operator, with $b^{\dagger}_p$($b_p$) and $d^{\dagger}_p$($d_p$) the fermionic and antifermionic creation(annihilation) mode operators with frequency $\omega_p$, being $a^{\dagger}_k$($a_k$) the creation(annihilation) bosonic mode operator associated with the frequency $\omega_k$. 

With the aim to adapt the model to be simulated in the simulating platform, we take into account a last simplification in Eq.~(\ref{complex}): (iii) a single fermionic and a single antifermionic modes~\cite{Casanova11,QFTKihwanKim} which will interact through a continuum of bosonic field modes. The latter, in the spirit of a digital-analog quantum simulator, is introduced to study a coupled field theory that may represent fermion-fermion collisions, dressed states, pair creation and annihilation, as well as non-perturbative scenarios.

We define now the $j$th input comoving fermionic and antifermionic modes in Schr\"{o}dinger picture, which read~\cite{Casanova11}
\begin{eqnarray}
b^{\dag (j)}_{\rm in} &=& \int dp \ \Omega^{(j)}_f(p^{(j)}_{f},p)b^{\dag}_pe^{-i\omega_p t} ,\label{comovingb} \\
d^{\dag (j)}_{\rm in} &=& \int dp \ \Omega^{(j)}_{\bar{f}}(p^{(j)}_{\bar{f}},p)d^{\dag}_pe^{-i\omega_p t} \label{comovingd} ,
\end{eqnarray}
where $\Omega^{(j)}_{f,\bar{f}}(p^{(j)}_{f,\bar{f}},p)$ denote the $j$th fermionic and antifermionic field mode wave packets with average momenta $p_f$ and $p_{\bar{f}}$, respectively. These particle modes will produce physical comoving wave packets when acting on the vacuum which are appropriate to describe physical particles, unlike the standard plane waves which are totally delocalized and not normalizable. For practical purposes, we will restrict to orthonormal wave packets $\Omega^{(j)}_{f,\bar{f}}(p^{(j)}_{f,\bar{f}},p)$, in such a way that the comoving fermionic and antifermionic modes satisfy the equal-time anti-commutation relations $\{ b^{(i)}_{\rm in},b_{\rm in}^{\dag (j)}\} = \delta_{ij}$.

The realization of the dynamics given by Hamiltonian~(\ref{complex}) in a circuit QED setup is a complex problem given that it describes an infinite amount of both fermionic and bosonic modes. In order to reproduce the latter, we will employ the bosonic-mode continuum which appears in open transmission lines as well as low-Q cavities. To be able to cope with the former, we simplify the fermionic field $\psi(x)$ as comprised of a discrete and truncated set of comoving fermionic modes. The latter condition will permit to express $\psi(x)$ as a superposition of two of these fermionic anticommuting modes as the lowest order of an expansion, and neglect the other anticommuting modes as not populated at lowest order. Therefore, the fermionic mode field is given as
\begin{equation}\label{coferm}
\psi(x) \simeq \Lambda_1(p^{(1)}_{f},x,t)b^{(1)}_{\rm in}+\Lambda_2(p^{(1)}_{\bar{f}},x,t)d^{\dag(1)}_{\rm in} ,
\end{equation}
where 
\begin{eqnarray}
\Lambda_1(p^{(1)}_{f},x,t) = \{\psi(x),b^{\dag(1)}_{\rm in}\}
 = \frac{1}{\sqrt{2\pi}} \int \frac{dp}{\sqrt{2\omega_{p}}} \Omega^{(1)}(p^{(1)}_{f},p)e^{i(px-\omega_p t)} ,\label{cb} \\ 
\Lambda_2(p^{(1)}_{\bar{f}},x,t) = \{\psi(x),d^{(1)}_{\rm in}\}
= \frac{1}{\sqrt{2\pi}} \int \frac{dp}{\sqrt{2\omega_{p}}} \Omega^{(1)}(p^{(1)}_{\bar{f}},p)e^{-i(px-\omega_p t)},\label{cd}\nonumber\\
\end{eqnarray}
and we consider $\psi(x)$ in the Schr\"{o}dinger picture. In the remainder of the section, we will omit the superindex $(1)$ given that we only employ these two creation operators.

Accordingly, we can now rewrite the interaction Hamiltonian associated with the introduced quantum field theory in the light of the previous considerations. We substitute the definitions for the bosonic $A(x)$ and fermionic $\psi(x)$ fields in the expression of Eq.~(\ref{complex}) to obtain 
\begin{eqnarray}\label{simple}
H_{\rm int} &=& i \int dx dk\lambda_k\sqrt{\frac{\omega_{k}}{2}} \ \Big( |\Lambda_1(p_{f},x,t)|^{2} b^\dag_{\rm in}b_{\rm in} 
 + \Lambda_1^*(p_{f},x,t)\Lambda_2(p_{\bar{f}},x,t) b^\dag_{\rm in}d^\dag_{\rm in}  \\ 
&& + \Lambda_2^*(p_{\bar{f}},x,t)\Lambda_1(p_{f},x,t) d_{\rm in}b_{\rm in}  + |\Lambda_2(p_{\bar{f}},x,t)|^{2} d_{\rm in}d^\dag_{\rm in}\Big) \left( a^{\dag}_k e^{-i k x} - a_k  e^{i k x} \right) \! .\nonumber
\end{eqnarray}
The fermionic and antifermionic field modes satisfy standard anticommutation relations, $\{b_{\rm in},b^\dag_{\rm in}\}=\{d_{\rm in},d^\dag_{\rm in}\} = 1$, while the bosonic field modes satisfy standard conmutation relations $[a_k,a^{\dag}_{k^{\prime}}]\!=\!\delta(k-k')$. Accordingly, one may expect that reproducing a model with a discrete set of fermionic modes, which interact with a full-fledged continuum of bosonic modes, will be a significant step forward in the implementation of quantum field theories with controllable quantum platforms. This is fully in the spirit of the digital-analog quantum simulator concept.

\begin{figure}[h]
	\centering
	\includegraphics[width=1\linewidth]{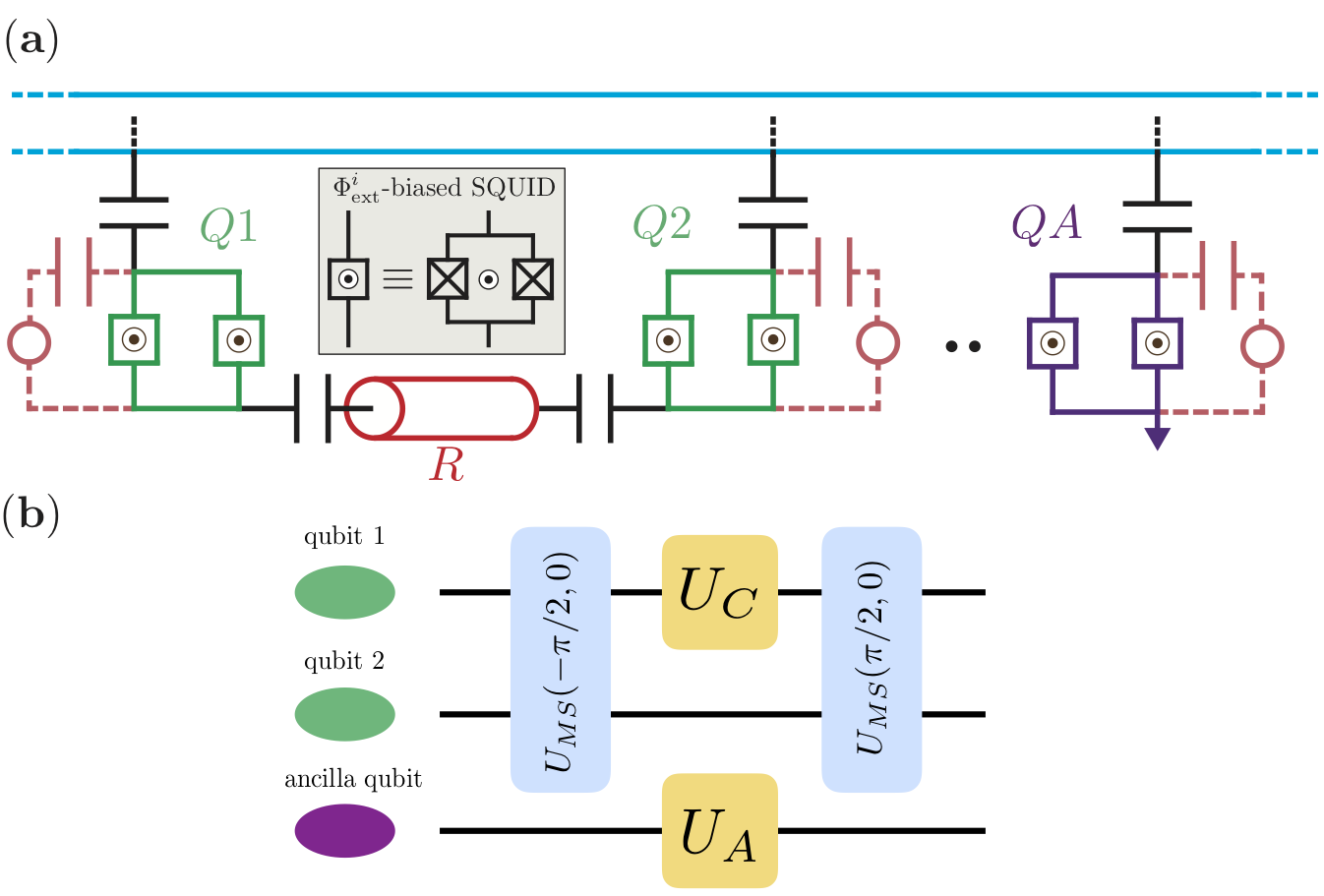}
	\caption{\label{fig:FFscattering_fig} \textbf{Proposal \cite{QFTcQED,GarciaAlvarez16} for simulating fermion-fermion scattering in quantum field theories.} (a) An open transmission line (blue) approximately supporting the continuum of bosonic modes interacts with two flux-tunable superconducting transmon qubits $Q1$ and $Q2$ (green), that simulate fermionic degrees of freedom, and one ancilla qubit (violet). A second waveguide (red), terminated in capacitors at each edge, sustains a single mode of the microwave field which also interacts with the qubits. Individual addressing of these qubits is permitted through flux lines to tune their coupling and self-energies at will (brown circles). (b) Single Trotter step sequence of single and multiqubit gates that generate two-qubit interactions coupled to the continuum. Here,  $U_{\text{MS}}(-\pi/2)=\exp(i\pi \sigma^x \otimes \sigma^x /4)$ and $U_{C,A}=\exp[-\phi\sigma_{1,A}^{z,y}\int dk g_k (a_k^\dagger \exp^{-ikx} -a_k \exp^{ikx})]$.}
\end{figure}

In order to implement the previous model in a circuit QED platform, one may consider to encode the fermionic modes onto qubits with the Jordan-Wigner transformation, and to couple them with the bosonic continuum via M{\o}lmer-S{\o}rensen (MS) and local gates as described in Sections \ref{sec:JW} and \ref{sec:MolmerSorensen}, as follows. A scheme of a possible implementation is depicted in Fig.~\ref{fig:FFscattering_fig}.a. To give an example, a plausible interaction term is $H = i( b^{\dag}_id^{\dag}_j +  d_jb_i)\int dk g_k (a_k^{\dag}e^{-ikx}-a_ke^{ikx} )$. The Jordan-Wigner transformation permits to express the above interaction Hamiltonian as the tensor product of Pauli operators coupled with a continuum interval of bosonic field modes. To achieve this, one may combine two multiqubit MS gates with a local gate coupling one of the qubits with the continuum of bosons~\cite{Casanova12,Mezzacapo12}, as shown in Fig.~\ref{fig:FFscattering_fig}.b,
\begin{eqnarray}
U \! & = & \! U_{\rm MS}(-\pi/2, 0)U_{\sigma_z}(\phi)U_{\rm MS}(\pi/2,0)\nonumber\\
&=&\exp{[\phi(\sigma^z\otimes\sigma^x \otimes \sigma^x\otimes ...) \int dk g_k(a_k^{\dag}e^{-ikx}-a_ke^{ikx} )]} , \!\!
\end{eqnarray}
where $U_{\rm MS}$ is the M\o lmer-S\o rensen operator~\cite{Molmer99} which may be expressed as $U_{\rm MS}(\theta,\phi)=\exp[-i\theta(\cos\phi S_x+\sin \phi S_y)^2/4]$. Here,  $S_{x,y}=\sum_{i}\sigma_i^{x,y}$ acts on as many qubits as fermionic field modes involved, while the local gate  $U_{\sigma_z}(\phi)$ is  $\exp[-\phi\sigma_1^z \int dk g_k (a_k^{\dag}e^{-ikx}-a_ke^{ikx} )]$. 

Circuit QED platforms which include the interaction between transmon qubits and on-chip open transmission lines~\cite{Gambetta2011,Houck2011,Koch07} are a suitable system to achieve the requirements of the introduced digital-analog quantum simulator. We consider a microwave open transmission line comprising a continuum of microwave field modes which interact with three transmon qubits, two of them implementing the fermionic and antifermionic modes, and the third one, an ancilla, in order to couple them to the continuum of bosonic modes. In addition, the protocol employs a microwave resonator with a single electromagnetic mode coupled only with the two transmons representing fermionic and antifermionic modes. This will allow for the coupling between them, for the pair creation and annihilation. We point out that two transmon qubits may be simultaneously coupled with both resonators, while the ancilla is only coupled to the open transmission line. 

In this protocol, tunable couplings between each qubit and the transmission lines will be needed, as well as tunable qubit energies through magnetic flux drivings. The implementation simulating fermion-fermion scattering will need the ability to switch on and off each qubit-resonator coupling via control parameters. These may be achieved by considering tunable coupling transmon qubits,~\cite{Gambetta2011,Houck2011} as well as standard band-stop filters~\cite{Pozar} on the coplanar open transmission line, such that only a bandwidth of bosonic field modes enters into the evolution. Finally, this approach may be boosted forward to include several fermionic field modes via the addition of further transmon qubits.

\section{Perspectives towards digital-analog quantum computing}\label{sec:DAQC}
The aim of this section is to discuss the role of the digital-analog approach as a novel quantum computing paradigm, and raise the questions which must be addressed to achieve this goal. Even though the perspectives of the digital-analog approach are promising, there is still a hard theoretical and experimental path to follow. Although the concepts of digital and analog quantum simulations are broadly used, they lack of a formal mathematical definition, and the frontiers between them are blurry. Therefore, if we want to formally discuss the digital-analog quantum computing (DAQC) paradigm, we need firstly to establish formally the border, and based on this, properly define the new approach. 

Once a formal definition of DAQC is obtained, a natural question is to wonder about the universality of the scheme, i.e. whether any unitary may be simulated within this framework. A property which is usually highlighted in DAQS is that the number of resources required in terms of number of entangling gates is dramatically reduced when compared with the purely digital approach. However, until now, while plausible, this is a purely empirical statement which, therefore, must be formally proven within a formal context of DAQC. Additionally, as a consequence of this reduction in the resources, it is normally shown that the error scales better for larger systems, but this is again an empirical result which is consequence of numerical and experimental simulations, but the question of {\it how} this scaling is remains open. Indeed, it is natural to think that this improvement strongly depends on the relation between the simulated dynamics (unitary) and the implementable analog blocks in the simulating platform.

Additional questions would be whether there exist systematic methods to efficiently decompose a given dynamics within a DAQS/DAQC approach, to face the possibility of designing quantum error correction protocols adapted to this framework, or how to combine it in a hybrid classical-quantum approach to increase the flexibility \cite{Unai2016,Moll2017}.

To sum up, based on the aforementioned issues, we consider that there is still a series of open questions in the avenue to formally establish a DAQC paradigm which deserves further work and attention.

\section{Conclusions}
In the field of quantum simulations, historically, either the digital or the analog approaches have been pursued. While the former provides flexibility and universality, the latter can achieve better scalability. In this review article, we have analyzed a third possibility that merges the previous two taking the best of each of them: a digital-analog quantum simulator. This kind of devices, employing digital steps in combination with analog blocks, and without quantum error correction protocols, will be able to simulate a variety of quantum complex systems in a scalable and versatile way, paving the way to reaching a quantum advantage with quantum technologies.

\section{Funding}

The authors acknowledge support from Spanish MINECO/FEDER FIS2015-69983-P, Ram\'on y Cajal Grant RYC-2012-11391, Basque Government IT986-16 and Ph.D. Grant No. PRE-2016-1-0284. This material is also based upon work supported by the U.S. Department of Energy, Office of Science, Office of Advance Scientific Computing Research (ASCR), under field work proposal number ERKJ335.

\end{document}